# NMR Based Diffusion Pore Imaging by Double Wave Vector Measurements

Tristan Anselm Kuder [1] and Frederik Bernd Laun [1,2]

[1] *Medical Physics in Radiology, German Cancer Research Center (DKFZ), Heidelberg, Germany*

[2] *Quantitative Imaging Based Disease Characterization, German Cancer Research Center (DKFZ), Heidelberg, Germany*

**Abstract**

In porous material research, one main interest of nuclear magnetic resonance (NMR) diffusion experiments is the determination of the exact shape of pores. It has been a longstanding question if this is achievable in principle. In this work, we present a method using short diffusion gradient pulses only, which is able to reveal the shape of arbitrary closed pores without relying on a priori knowledge. In comparison to former approaches, the method has reduced demands on relaxation times and allows for a more flexible NMR sequence design, since, for example, stimulated echoes can be used.

Nuclear magnetic resonance (NMR) measurements are extensively used as non-invasive imaging technique to investigate porous media and biological tissues [1, 2]. To gain information about the structure of a porous medium, a high spatial resolution of the acquired NMR images is essential. In conventional NMR imaging, magnetic field gradients **G** are superimposed on the static magnetic field $B_0$ yielding a spatial dependence of the Larmor frequency $\omega = \gamma(B_0 + \mathbf{x} \cdot \mathbf{G})$ on the location **x** of the spins, which finally allows to reconstruct an image of the spin distribution in the sample using a Fourier transform (FT) [3]. With increasing gradient strength and better resolution, the size of the image volume elements is reduced and consequently the detectable signal decreases. Therefore, the achievable resolution is not only limited by the maximum gradient strength but also by the remaining signal-to-noise ratio (SNR).

In many porous systems however, the main interest of NMR measurements is the determination of the shape of the pores, and not of their position in the sample. A widely used approach in this regard is the NMR based detection of the diffusion process inside the pores. This allows inferring information on the pore structure which hinders the diffusion of the spins, while the signal can be collected from a volume much larger than the pore size [1, 2, 4-9]. Although much effort has been put in this area of research, it as been an open question for a long time, if it is possible to reconstruct the shape of closed pores from the diffusion weighted NMR signal [10].

Recently, two methods have been proposed to obtain the exact pore shape by NMR diffusion measurements [11, 12]. Since all pores can contribute to the signal, an "average pore" can be measured and the above mentioned SNR limit of conventional NMR imaging is lifted. The first method published in our previous work [11] uses a combination of a very long and a short diffusion weighting gradient pulse. The second method [12], which is considered here, relies on the application of short gradient pulses, which may be advantageous regarding the flexibility of the measurement sequence design. In the current work, we show that this second approach, as it was initially described in [12], is of limited use since it can only be applied to point symmetric domains. We show that the data acquired using the second approach [12] can nonetheless be used to reconstruct the shape of arbitrary domains. Moreover, we demonstrate that the data acquisition can, in principle, be simplified considerably, and compare the new approach to the first approach described in [11]. These findings render the diffusion pore imaging technique applicable to a much wider class of pore types and sizes, since it allows one to use NMR measurement sequences that rely on stimulated echoes, and since it shows better convergence properties.

For the NMR based detection of diffusing molecules, a gradient profile $\mathbf{G}(t)$ depending on the time $t$ is applied fulfilling the rephasing condition $\int dt\,\mathbf{G}(t) = 0$. A random walker following the trajectory $\mathbf{x}(t)$ acquires the phase $\phi = -\gamma \int_0^T dt\,\mathbf{G}(t)\cdot\mathbf{x}(t)$ and the signal attenuation due to the resulting phase dispersion is $S = \langle \exp(i\phi)\rangle$, where $\langle \cdot \rangle$ denotes the average over all possible paths and $T$ the total duration of the gradient profile. For resting particles, the NMR signal is not affected by the diffusion gradients. A particularly well known technique is q-space imaging [7], which uses two short gradient pulses of duration $\delta$ and gradient vectors $\mathbf{G}_1 = \mathbf{G}$ and $\mathbf{G}_2 = -\mathbf{G}$. It is assumed that $\delta$ is so short that the diffusive motion can be neglected during the gradient duration and that $T$ is sufficiently long for the spins to explore the whole pore. Thus, the correlations of the starting point $\mathbf{x}_1$ (the particle location during the application of $\mathbf{G}_1$) and the final point $\mathbf{x}_2$ (application of $\mathbf{G}_2$) of the random walker are lost. Introducing $\mathbf{q} = \gamma\delta\mathbf{G}$ and the pore space function $\rho(\mathbf{x})$, which equals zero outside and the reciprocal pore volume inside the pore, the signal attenuation $S$ becomes for this gradient profile

$$S_2(\mathbf{q}) = \langle e^{i\mathbf{q}\cdot(\mathbf{x}_2-\mathbf{x}_1)}\rangle = \langle e^{i\mathbf{q}\cdot\mathbf{x}_2}\rangle\langle e^{-i\mathbf{q}\cdot\mathbf{x}_1}\rangle = \int_V d\mathbf{x}_2\,\rho(\mathbf{x}_2)e^{i\mathbf{q}\cdot\mathbf{x}_2}\int_V d\mathbf{x}_1\,\rho(\mathbf{x}_1)e^{-i\mathbf{q}\cdot\mathbf{x}_1} = |\tilde{\rho}(\mathbf{q})|^2. \quad [1]$$

Here, $\tilde{\rho}(\mathbf{q})$ denotes the Fourier transform of $\rho(\mathbf{x})$ and $V$ is the volume, in which the pore is located. Using this approach, only the power spectrum $|\tilde{\rho}(\mathbf{q})|^2$ of the pore space function can be measured. Due to the lost phase information, $\rho(\mathbf{x})$ cannot be reconstructed from $S_2(\mathbf{q})$.

To preserve the phase information, the method described in [12] uses an additional measurement with three equally spaced gradient pulses of duration $\delta$ with the gradient vectors $\mathbf{G}$, $-2\mathbf{G}$ and $\mathbf{G}$. The total diffusion time $T$ is so large, that the long time limit is reached in each of the two time intervals between the gradients. Similarly to Eq. 1, one finds for the signal attenuation

$$S_3(\mathbf{q}) = \tilde{\rho}(\mathbf{q})^2 \, \tilde{\rho}^*(2\mathbf{q}). \qquad [2]$$

Contrary to $S_2(\mathbf{q})$, $S_3(\mathbf{q})$ contains phase information, but the desired phase of $\tilde{\rho}(\mathbf{q})$ is in general not directly accessible. Only for point symmetric domains $(\rho(\mathbf{x}) = \rho(-\mathbf{x}))$, there is a simple solution, since $\tilde{\rho}(\mathbf{q})$ is real if and only if the pore is point symmetric. To prove this, we define $\rho_2(\mathbf{x}) = \rho(-\mathbf{x})$ and, keeping in mind that $\rho(\mathbf{x})$ is real, obtain for the FT of $\rho_2(\mathbf{x})$

$$\tilde{\rho}_2(\mathbf{q}) = \int_V d\mathbf{x}\, \rho_2(\mathbf{x}) e^{-i\mathbf{q}\mathbf{x}} = \int_V d\mathbf{x}'\, \rho_2(-\mathbf{x}') e^{i\mathbf{q}\mathbf{x}'} = \int_V d\mathbf{x}'\, \rho(\mathbf{x}') e^{i\mathbf{q}\mathbf{x}'} = \tilde{\rho}^*(\mathbf{q}). \qquad [3]$$

$\rho(\mathbf{x})$ is point symmetric if and only if $\rho_2(\mathbf{x}) = \rho(\mathbf{x})$ for all $\mathbf{x}$, which is equivalent to $\tilde{\rho}_2(\mathbf{q}) = \tilde{\rho}(\mathbf{q})$ for all $\mathbf{q}$. According to Eq. 3 this is equivalent to $\tilde{\rho}^*(\mathbf{q}) = \tilde{\rho}(\mathbf{q})$ for all $\mathbf{q}$.

Thus, for point symmetric domains, $\rho(\mathbf{x})$ can be directly obtained from the quotient of the two measurements by FT (**method 1**)

$$N(\mathbf{q}) = \frac{S_3(\mathbf{q})}{S_2(\mathbf{q})} = \frac{\tilde{\rho}(\mathbf{q})^2 \, \tilde{\rho}^*(2\mathbf{q})}{|\tilde{\rho}(\mathbf{q})|^2} = \tilde{\rho}^*(2\mathbf{q}), \qquad [4]$$

However, for arbitrary pore shapes, $\tilde{\rho}(\mathbf{q})$ is not real and the last part of the equation is not valid, which is not mentioned in [12].

Fig. 1(a) shows $S_2(\mathbf{q})$, $S_3(\mathbf{q})$ and $N(\mathbf{q}/2)$ for a cylindrical pore, for which $\tilde{\rho}(\mathbf{q})$ is real and for which Eq. 4 is valid. The calculation was performed analytically in the limit of long diffusion time and $\delta \to 0$. Due to the point symmetry, the curves for $N(\mathbf{q}/2)$ and $\tilde{\rho}^*(\mathbf{q})$ exactly coincide. For an equilateral triangle, $\tilde{\rho}(\mathbf{q})$ is in general complex, which is shown for the vertical gradient direction in Fig. 1(b). Eq. 4 is no longer valid, all the zero crossings of $\tilde{\rho}(\mathbf{q})$ are lost in $N(\mathbf{q})$, and the approach described in [12] is not applicable.

For arbitrary domains, we propose an iterative approach to recover the phase information from $S_3(\mathbf{q})$. We write $S_3(\mathbf{q})$ and $\tilde{\rho}(\mathbf{q})$ in terms of magnitude and phase as

$$S_3(\mathbf{q}) = M(\mathbf{q}) \cdot e^{i \cdot \varphi(\mathbf{q})} \text{ and } \tilde{\rho}(\mathbf{q}) = A(\mathbf{q}) \cdot e^{i \cdot \psi(\mathbf{q})}. \qquad [5]$$

Therefore, as $A(\mathbf{q})$ can readily be computed from $S_2(\mathbf{q})$ by $A(\mathbf{q}) = \sqrt{S_2(\mathbf{q})}$, and using Eq. 2, we obtain a recursive equation for the unknown phase

$$\psi(\mathbf{q}) = 2\psi(\mathbf{q}/2) - \varphi(\mathbf{q}/2). \qquad [6]$$

It is assumed that $S_2(q\mathbf{n})$ and $S_3(q\mathbf{n})$ are measured for different values $q$ for a specific gradient direction $\mathbf{n}$. As Eq. 6 must be solved recursively, appropriate initial values must be specified. Due to the rephasing condition, $S_2(\mathbf{q})$ and $S_3(\mathbf{q})$ do not change if a pore is shifted to another position. This, in turn, leaves one the freedom to chose an arbitrary position of the pore in the reconstructed image. A natural choice is to set the center of mass of the pore to the origin. Then, when expanding $\tilde{\rho}(\mathbf{q})$ in $\mathbf{q}$, it follows that the term linear in $\mathbf{q}$ vanishes

$$\int_V d\mathbf{x}\rho(\mathbf{x})e^{-i\mathbf{q}\cdot\mathbf{x}} = \int_V d\mathbf{x}\rho(\mathbf{x})\left(1 - i\mathbf{q}\cdot\mathbf{x} + O(\mathbf{q}^2)\right) = 1 + 0 + O(\mathbf{q}^2), \qquad [7]$$

Thus, $\psi(0) = 0$ and $\partial\psi(\mathbf{n}\cdot q)/\partial q\,|_{q=0} = 0$ are the corresponding initial conditions. If the reconstructed pore image shall be shifted by $\Delta\mathbf{x}$, the initial values are specified by $\psi(0) = 0$ and $\partial\psi(\mathbf{n}\cdot q)/\partial q\,|_{q=0} = -\Delta\mathbf{x}$. Assuming $\psi(\mathbf{n}\,q_1/2) = 0$ for the smallest non-zero $q$-value $q_1$, $\psi(q\mathbf{n})$ can be iteratively estimated with appropriate interpolation, provided that the radial acquisition contains sufficient data points. Thus, as $A(q\mathbf{n})$ and $\psi(q\mathbf{n})$ are known, $\tilde{\rho}(q\mathbf{n})$ is determined uniquely (we label this approach **method 2**). Fig. 1(c) shows the phases $\varphi(q\mathbf{n})$ and $\psi(q\mathbf{n})$ for the triangular domain and the phase that was estimated by this iterative approach for 100 points, which are in excellent agreement with the exact curve.

We show now, that the q-space measurement of $S_2(\mathbf{q})$ can be omitted, since the magnitude $A(\mathbf{q})$ can be inferred from $S_3(\mathbf{q})$. Eq. 2 yields

$$A(\mathbf{q}) = \frac{M(\mathbf{q}/2)}{A(\mathbf{q}/2)^2}. \qquad [8]$$

With the condition $1 - A(q\mathbf{n}) \propto q^2$ (Eq. 7) for small $q$, $M(q\mathbf{n})$ obtained from radial acquisitions of $S_3(q\mathbf{n})$ can be used for the iterative estimation of $A(q\mathbf{n})$ and finally of $\tilde{\rho}(q\mathbf{n})$ (**method 3**).

If it is known a priori that the domain is point symmetric, method 1 can be applied as mentioned above. However, due to the quotient in Eq. 4, very large errors can result from minor noise contributions, if $S_2(\mathbf{q})$ is small. These errors can be easily reduced, because $\tilde{\rho}(\mathbf{q})$ can be extracted from $S_2(\mathbf{q})$ and from the sign of $S_3(\mathbf{q})$, which is real for point symmetric domains (**method 4**)

$$\tilde{\rho}(\mathbf{q}) = \sqrt{S_2(\mathbf{q})}\,\text{sign}(S_3(\mathbf{q}/2)). \tag{9}$$

For comparison, we recall the first approach for diffusion pore imaging presented in [11]. Here, a temporarily asymmetric gradient profile is used given by $\mathbf{G}(t) = -\delta(T-\delta)^{-1}\mathbf{G}$ for $0 \leq t \leq (T-\delta)$ and $\mathbf{G}(t) = \mathbf{G}$ for $(T-\delta) < t \leq T$. During the first long gradient of small amplitude, a diffusing particle acquires a phase that is equal to the phase a particle resting at the center of mass $\mathbf{x}_{\text{cm},1}$ of the random walk would get. The second gradient with the short duration $\delta$ yields a phase depending on the final particle position $\mathbf{x}_2$ and acts as an imaging gradient. If $T$ is sufficiently long for the diffusion process to reach the long time limit, $\mathbf{x}_{\text{cm},1}$ converges to the center of mass of the pore $\mathbf{x}_{\text{cm}}$. Thus, the signal attenuation $S_D(\mathbf{q})$ is

$$S_D(\mathbf{q}) = \left\langle e^{i\mathbf{q}\cdot(\mathbf{x}_{\text{cm}}-\mathbf{x}_2)} \right\rangle = e^{i\mathbf{q}\cdot\mathbf{x}_{\text{cm}}} \left\langle e^{-i\mathbf{q}\cdot\mathbf{x}_2} \right\rangle = e^{i\mathbf{q}\cdot\mathbf{x}_{\text{cm}}} \tilde{\rho}(\mathbf{q}) \tag{10}$$

and $\tilde{\rho}(\mathbf{q})$ can be directly measured (**method 5**). The prefactor $e^{i\mathbf{q}\cdot\mathbf{x}_{\text{cm}}}$ shifts the center of mass of all pores to a single point, enabling the measurement of an averaged pore space function.

To validate these approaches with finite diffusion time $T$ and gradient amplitude, we used the matrix approach described in [10, 13-15] which is similar to the approach in [16, 17] and allows an efficient and accurate simulation of the diffusion process for arbitrary gradient profiles with the help of precalculated eigenfunctions of the Laplace operator for the respective domain. For this purpose, a cylindrical and an equilateral triangular domain were used with the following parameters: diameter of the cylinder and edge length of the triangle $L = 25\,\mu\text{m}$, total duration of the gradient profiles $T = 1\,\text{s}$, $\delta = 4\,\text{ms}$, maximum $q$-value $q_{max} = 4.19\,\mu\text{m}^{-1}$ yielding a resolution of 0.75 µm and a maximum gradient strength of $3.9\,\text{Tm}^{-1}$. The signals were calculated for 90 gradient directions and 100 different $q$-values. To obtain the images from the projections, an inverse Radon transform was applied. Fig. 2 shows the results for the different methods without noise, while the influence of noise is displayed in Fig. 3.

Method 1, which is directly applying the FT to $N(\mathbf{q})$, is successful for the cylinder, while it completely fails for the triangle due to the lost phase information (Figs. 2(a,b)). When using the phase estimation (method 2, Figs. 2(c,d)), both domains can be imaged in good quality due to the reconstructed phase information. Near the edges of the domains, especially in the corners of the triangle, the edge enhancement effect [18-20] can be observed. Due to the diffusive motion during finite gradient pulse time $\delta$, the reconstructed size is slightly reduced and the edges appear brighter. The pore space function can even be reconstructed using solely $S_3(\mathbf{q})$ (Figs. 2(e,f); method 3). Small artifacts in (Fig. 2(e)) result from the amplification of numerical errors in the iterative estimation of $A(\mathbf{q})$, that are especially occurring at small values $A(\mathbf{q}/2)^2$ in Eq. 8. To reduce these effects, outliers were suppressed and replaced by interpolated values in Figs. 2(e,f). As expected, method 4 only utilizing the sign of $S_3(\mathbf{q})$ is successful for the cylinder, while it fails for the triangle, since $\tilde{\rho}(\mathbf{q})$ has imaginary parts (Figs. 2(g,h)). Figs. 2(i,j) depict the results using asymmetric gradients for comparison.

For the calculation of the images in Fig. 3, Gaussian noise with the standard deviation $\sigma = SNR^{-1}$ with $SNR = 1000$ was added to the real and the imaginary part of the signal; all other parameters were identical to Fig. 2. While method 1 completely fails due to the division of noisy data (Figs. 3(a,b)), method 2 and 3 allow a reasonable reconstruction of the cylinder and the triangle (Figs. 3(c-f)). As expected, method 3 (Figs. 3(e,f)) is more sensitive to noise than method 2. The sign measurement (method 4) is slightly more stable (Fig. 3(g)) than the phase estimation. Interestingly, the approach using asymmetric gradients (method 5, Figs. 3(i,j)) is practically unaffected by the added noise. The reason for the largely increased ro-bustness of method 5 can be understood looking at Fig. 1(a). The magnitude of $S_2(\mathbf{q})$ and $S_3(\mathbf{q})$ drops much faster with increasing $q = |\mathbf{q}|$ than $\tilde{\rho}(\mathbf{q})$. Therefore, the oscillations of $\tilde{\rho}(\mathbf{q})$ at higher $q$-values are contained in the very small values of $S_2(\mathbf{q})$ and $S_3(\mathbf{q})$ and cannot be reconstructed even if relatively small contributions of noise are present. Method 5, however, can directly measure $\tilde{\rho}(\mathbf{q})$ and resolve the oscillations at higher $q$-values. The reconstruction of the cylinder seems to be more problematic than that of the triangle using method 2 and 3, because, for the cylinder, more small values $A(\mathbf{q})$ occur, for which method 2 and 3 tend to amplify noise contributions. More advanced reconstruction methods, for example employing regularization techniques, would probably be able to increase the stability of these methods.

Looking at Fig. 2, a considerable smearing is observable when using method 5 (Figs. 2(i,j)), which is not present when using the other methods. For method 5, the convergence of the center of mass of the random walk to the pore center of mass $\mathbf{x}_{cm}$ is very slow when increasing $T$. The effect of better convergence properties for the methods using short gradients is much more pronounced for shorter $T$ yielding sharper images.

In summary, we have shown that it is possible to infer the exact shape of arbitrary closed pores using short gradient pulses only. In contrast to method 5, stimulated echoes could be used which may be advantageous to reach the long time limit, since the magnetization could be stored in the longitudinal direction between the gradient pulses taking advantage of the larger longitudinal relaxation time $T_1$ compared to the transversal one ($T_2$). Due to this fact and due to the better convergence properties, the methods presented here using short gradients only may be in some cases superior to the initial approach using asymmetric gradients [11]. Considering the fact that *SNR* values larger than 1000 seem to be reachable in practical applications according to the data presented in [12] and that extremely high gradient amplitudes are available [21], the diffusion pore imaging approach presented here may be a powerful tool for the investigation of porous media eliminating the fundamental limitation of conventional NMR imaging by the signal loss when the resolution is increased.

**Figures**

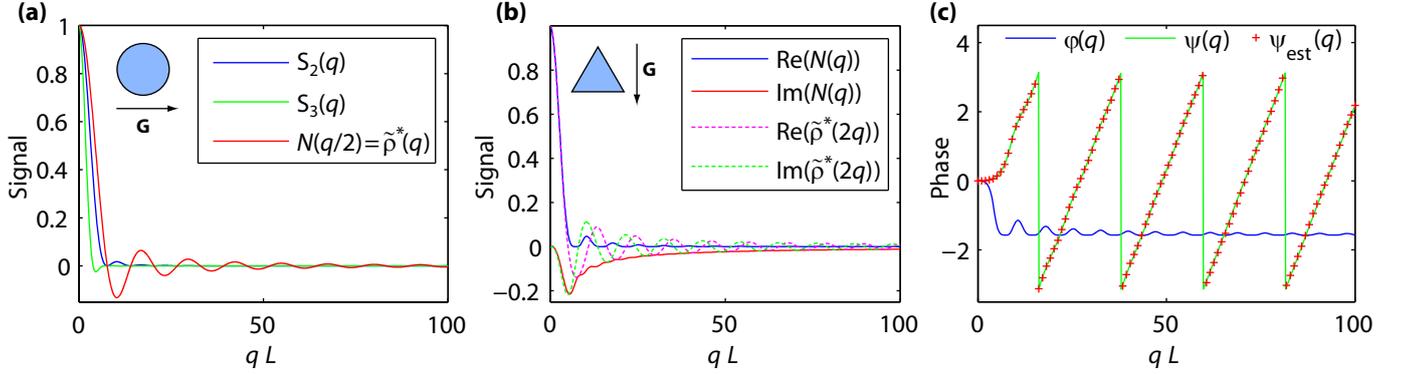

**Fig. 1**: Analytically calculated signal for a cylindrical and a triangular pore for one gradient direction $(q=|\mathbf{q}|)$. (a) $S_2(\mathbf{q})=|\tilde{\rho}(\mathbf{q})|^2$, $S_3(\mathbf{q})=\tilde{\rho}(\mathbf{q})^2\tilde{\rho}^*(2\mathbf{q})$ and $N(\mathbf{q}/2)=S_3(\mathbf{q}/2)/S_2(\mathbf{q}/2)$ for the cylindrical pore. Due to the point symmetry, $N(\mathbf{q})=\tilde{\rho}^*(2\mathbf{q})=\tilde{\rho}(2\mathbf{q})$ is valid. (b) Real and imaginary part of $N(\mathbf{q})$ and $\tilde{\rho}^*(2\mathbf{q})$ for an equilateral triangle and vertical gradient direction. Since $\rho(\mathbf{x})\neq\rho(-\mathbf{x})$, $N(\mathbf{q})\neq\tilde{\rho}^*(2\mathbf{q})$. (c) Phase $\varphi(\mathbf{q})$ of $S_3(\mathbf{q})$ and $\psi(\mathbf{q})$ of $\tilde{\rho}(\mathbf{q})$ as well as the phase $\psi_{est}(\mathbf{q})$ estimated for 100 measured points using Eq. 6. $L$ denotes the diameter of the cylinder and the edge length of the triangle.

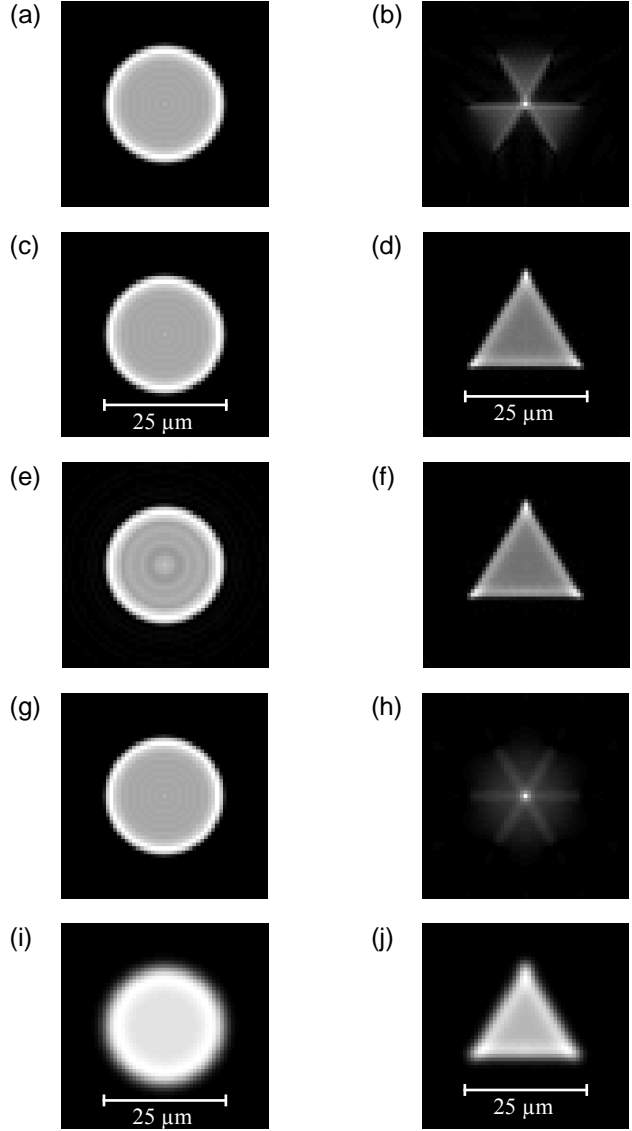

**Fig. 2**: Images reconstructed for a cylindrical (left column) and a triangular (right column) pore using method 1 to 5 (row 1 – 5). Simulations were performed using a matrix approach calculating the signal with the help of precalculated Laplacian eigenfuntions ($L = 25\,\mu m$, $T = 1\,s$, $\delta = 4\,ms$, $q_{max} = 4.19\,\mu m^{-1}$, resolution of 0.75 μm). While all methods are successful for the point symmetric cylinder, the triangle can only be obtained when preserving the correct phase information, which method 1 and 4 (b, h) disregard. Method 5 (i, j) directly measuring $\tilde{\rho}(\mathbf{q})$ has a slower convergence to the long time limit than the methods employing short gradients which results in a smearing of the borders.

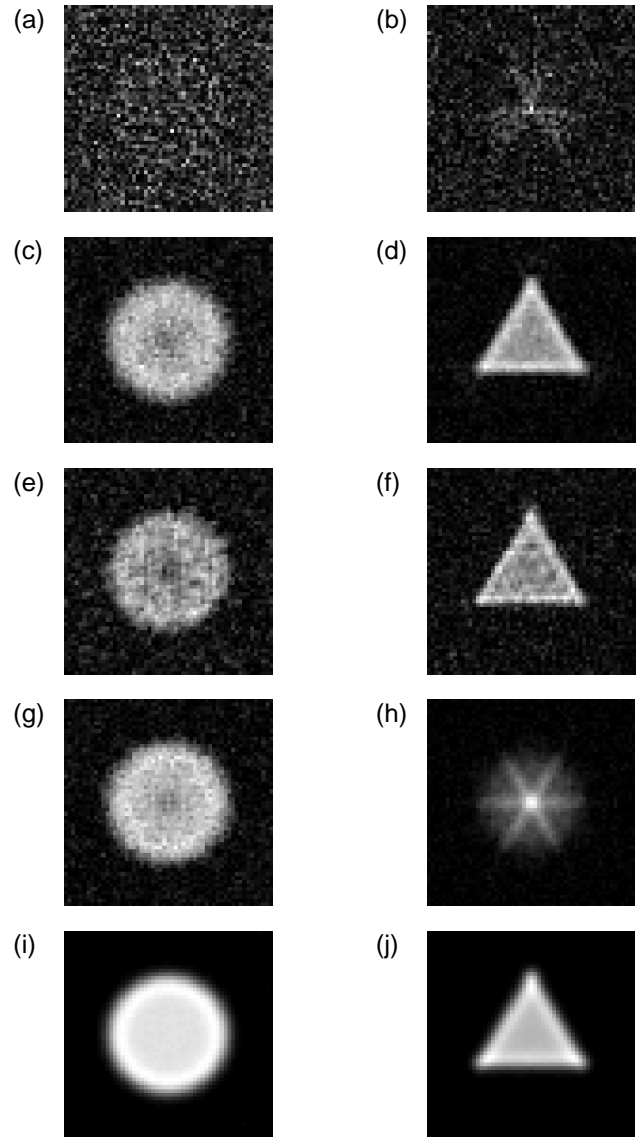

**Fig. 3**: Influence of noise on the reconstructed images; identical simulation as in Fig. 2 (rows: method 1 to 5), except for the addition of Gaussian noise with $\sigma = SNR^{-1}$ to the real and the imaginary part of the calculated signals $(SNR = 1000)$. For all methods calculated values $\tilde{\rho}(\mathbf{q}) > 1$ were set to zero. While method 1 is very unstable, methods 2-4 yield good results; however, method 5 is even more robust concerning noise.